\documentclass[a4paper,fleqn]{cas-dc}
\usepackage{url}  
\usepackage{graphicx} 
\usepackage[numbers]{natbib}  
\usepackage{caption} 
\usepackage{enumitem}
\usepackage{amsmath,amssymb,amsfonts}
\usepackage{algorithm, algorithmicx, algpseudocode}
\usepackage{textcomp}
\usepackage{xcolor}
\usepackage{tikz}
\usepackage{pgfplots}
\usepackage{subcaption}
\usepackage{multirow}
\usepackage{array}
\usepackage{booktabs}
\usepackage[mathcal]{eucal}
\usepackage{newfloat}
\usepackage{listings}
\usepackage[beginLComment=/*~,endLComment=~*/]{algpseudocodex}
\DeclareCaptionStyle{ruled}{labelfont=normalfont,labelsep=colon,strut=off} 
\floatstyle{ruled}
\newfloat{listing}{tb}{lst}{}
\floatname{listing}{Listing}
\usetikzlibrary{arrows.meta, positioning, shapes.geometric, fit, backgrounds, calc, decorations.pathreplacing}
\pgfplotsset{compat=1.17}

\definecolor{tier1color}{RGB}{52, 152, 219}
\definecolor{tier2color}{RGB}{155, 89, 182}
\definecolor{tier3color}{RGB}{231, 76, 60}
\definecolor{agentcolor}{RGB}{46, 204, 113}
\definecolor{clustercolor}{RGB}{241, 196, 15}

\newcommand{\hacn}{$\mathcal{HACN}$ }
\newcommand{\bigO}{\mathcal{O}}

\def\BibTeX{{\rm B\kern-.05em{\sc i\kern-.025em b}\kern-.08em
    T\kern-.1667em\lower.7ex\hbox{E}\kern-.125emX}}

\makeatletter
\newcounter{phase}[algorithm]
\newlength{\phaserulewidth}
\newcommand{\setphaserulewidth}{\setlength{\phaserulewidth}}
\newcommand{\phase}[1]{%
  \vspace{-1.25ex}
  \Statex\leavevmode\llap{\rule{\dimexpr\labelwidth+\labelsep}{\phaserulewidth}}\rule{\linewidth}{\phaserulewidth}
  \Statex\strut\refstepcounter{phase}\textit{\textbf{\color{purple}Phase~\thephase~--~#1}}
  \vspace{-1.25ex}\Statex\leavevmode\llap{\rule{\dimexpr\labelwidth+\labelsep}{\phaserulewidth}}\rule{\linewidth}{\phaserulewidth}}
\makeatother
\setphaserulewidth{.7pt}

\begin{document}
\let\WriteBookmarks\relax
\def\floatpagepagefraction{1}
\def\textpagefraction{.001}
\shorttitle{Consensus in Collaborative Multi-Agent AI Systems}
\shortauthors{Shit \& Subudhi}

\title [mode = title]{Hierarchical Adaptive Consensus Network: A Dynamic Framework for Scalable Consensus in Collaborative Multi-Agent AI Systems}

\author[1]{Rathin Chandra Shit}[
                        orcid=0000-0003-0642-9695]
\ead{rathin088@gmail.com}

\credit{Conceptualization of this study, Methodology, Writing-Original Draft Preparation}
\affiliation[1]{organization={Department of CSE, International Institute of Information Technology},city={Bhubaneswar}, postcode={751003},state={Odisha},country={India}}

\author[2]{Sharmila Subudhi}[
                        orcid=0000-0002-5043-1533]
\cormark[1]
\ead{sharmilasubudhi@ieee.org}

\credit{Software, Validation, Verification, Writing-Final Draft Preparation}
\affiliation[2]{organization={Department of CS, Maharaja Sriramm Chandra Bhanja Deo University},city={Baripada}, postcode={757003},state={Odisha},country={India}}
\cortext[cor1]{Corresponding author}

\begin{abstract}
The consensus strategies used in collaborative multi-agent systems (MAS) face notable challenges related to adaptability, scalability, and convergence certainties. These approaches, including structured workflows, debate models, and iterative voting, often lead to communication bottlenecks, stringent decision-making processes, and delayed responses in solving complex and evolving tasks. This article introduces a three-tier architecture, the Hierarchical Adaptive Consensus Network (\hacn), which suggests various consensus policies based on task characterization and agent performance metrics. The first layer collects the confidence-based voting outcomes of several local agent clusters. In contrast, the second level facilitates inter-cluster communication through cross-clustered partial knowledge sharing and dynamic timeouts. The third layer provides system-wide coordination and final arbitration by employing a global orchestration framework with adaptable decision rules. The proposed model achieves $\bigO(n)$ communication complexity, as opposed to the $\bigO(n^2)$ complexity of the existing fully connected MAS. Experiments performed in a simulated environment yielded a 99.9\% reduction in communication overhead during consensus convergence. Furthermore, the proposed approach ensures consensus convergence through hierarchical escalation and dynamic adaptation for a wide variety of complicated tasks.
\end{abstract}


\maketitle

\section{Introduction}\label{sec1}
The emergence of Multi-Agent Systems (MAS) with Large Language Models (LLM) brings opportunities that are unprecedented in terms of collaborative problem solving across a wide variety of domains \cite{li2024survey, NEURIPS2024_1c2b1c8f, Shit2018a}. Nevertheless, the realization of successful consensus between multiple AI agents is still a primary challenge that has a direct effect on the system reliability, efficiency, and scalability \cite{NEURIPS2024_ee71a4b1}. 

Current frameworks show three major consensus techniques: i) structured workflow-based methods, which allocate predetermined roles and sequential steps \cite{huang2025comfygpt}, ii) iterative voting-based methods where unanimity or a majority of the involved agents is necessary \cite{bauer2024multi, Shit2018b}, and iii) debate-based methods that rely on argumentation to refine the solution \cite{tran2025multiagent}. 

Structured approach, like MetaGPT \cite{hong2024metagpt} provides human-like organizational structures within the MAS environment by offering a task decomposition and error elimination through role specialization as Product Manager, Architect, Engineer, and QA Engineer. The Engineer agent conducts verification functions using Product Requirements Documents and system designs, while Error correction is achievable using an autonomous iterative workflow by role-checking outputs. Its strict sequential nature, however, introduces bottlenecks as well as single points of failure.

Iterative voting-based MAS provide explicit consensus protocols that can encounter stall-prone situations and undergo exponential communication complexity \cite{altermatt2025performance}. Such a method performs well in applying domain-specific tasks, like MedAgents \cite{tang_medagents}, because it can take advantage of expert knowledge and offer clear correction facilities to enhance accuracy.

Debate systems take advantage of divergent thinking and argumentation to make reasoning better \cite{liu2024groupdebate}. These frameworks solve degeneration-of-thought issues in which single agents could not produce new solutions anymore. Debate rules are generally organized as a series of rounds where an organized argument is required, with judge agents often set up to manipulate the debate and move towards better solutions. They do not often provide convergence guarantees and can be prone to circular arguments.

Although the solutions proposed in these approaches have unique benefits, all of them share common problems of congestion during scalability, instability in convergence, and the inability to adjust to dynamic needs in operation. We have highlighted a few limitations of these existing solutions that hinder their practicality when used in mass-scale implementations.

\begin{itemize}[label=-]
\item No formal convergence is guaranteed in most existing frameworks. Iterative voting systems can lead to stalemates when agents are in an irresolvable dispute, whereas debate-based methods can fall into circular reasoning with no assurance of a solution.
\item Fully connected models having $\bigO(n^2)$ overhead communication are impractical in a large-scale MAS ecosystem, restricting parallelization and scalability.
\item Existing frameworks exhibit restrictions in dynamically adapting to task attributes, time restrictions, or variation in flow agent performance, and adjusting the consensus mechanism \cite{shit2025cl}. This inflexibility often lowers performance in a wide range of operating conditions and does not allow to achieve optimal effort.
\item Existing models show inadequacy in attaining the Quality-Speed trade-off while trading consensus accuracy with a certain time limit. Systems are either thorough and slow or less detailed and fast to make decisions, with no adaptive process to best balance this trade-off.
\item Integrating heterogeneous agents having different abilities, knowledge scope, or trustworthiness presents a problem to the current MAS frameworks. 
\end{itemize}
    
In the light of these limitations, the current research elaborates a new three-layered architecture, called Hierarchical Adaptive Consensus Network (\hacn), to dynamically select a consensus strategy according to task properties as well as performance indicators of agents. The novelties of this approach are: 
\begin{enumerate}[label=(\roman*)]
\item A hierarchical scalable structure that achieves a reduced communication complexity from $\bigO(n^2)$ to $\bigO(n)$,
\item Adaptive decision rules that trade off between the quality of consensus development times, 
\item Use of voting mechanisms weighted by the confidence of a given voter, and
\item A hierarchical escalation-based algorithm that guarantees convergence.
\end{enumerate}
The rest of the paper is structured in the following way: Section \ref{sec3} discusses the design principles. Section \ref{sec4} describes the proposed \hacn architecture. Section \ref{sec5} focuses on the evaluation strategy, while Section \ref{sec6} depicts the implementation consideration. Section \ref{sec7} establishes the theoretical superiority in achieving convergence and scalability. Section \ref{sec8} presents the conclusion with implications and directions of future work.
\section{Suggested Framework Design}\label{sec3}
The \hacn schema is based on four key design principles addressing the needs of the existing consensus mechanisms.
\begin{enumerate}[label=(\alph*)]
\item \textbf{Hierarchical Decomposition for Scalability:} \hacn achieves scalability through a three-level hierarchy that distributes consensus decision across local clusters, inter-cluster coordination, and global orchestration. This breakdown simplifies the communication and does not compromise the quality of the decisions in terms of structured escalation.

\item \textbf{Adaptive Strategy Selection:} \hacn adopts a dynamic strategy selection mechanism by acclimatizing to the change in consensus strategies. Factors, such as the nature of the task, completion time of a task, and agents' performance measures, are used for determining a threshold value. The initial threshold is set as per the cluster number, iteration number, and solution list. In each iteration, 5\% threshold reduction is done to facilitate convergence. This flexibility allows the system to conform to decision quality as per the operating constraints and puts more focus on important decisions.

\item \textbf{Confidence-Weighted Decision Making for Reliability:} The solution confidence and historical agent accuracy are selected as critical factors in decision-making to enable participation of reliable agents only. The vote weight is computed by multiplying both the confidence scores and past performance measures. By doing so, we ascertain that the overconfident agents do not overwhelm the underperforming ones through unreliable voting. This voting structure enhances the solutions with better convergence properties.

\item \textbf{Graceful Degradation Under Constraints:} \hacn has several fallback components to impose a hierarchical escalation process when a consensus is not met at the lower level. This is done to do a quality control in such a way that the system behavior remains functional in the presence of hostile conditions, such as network partitions and time constraints. The hierarchical escalation retains the decision quality through structured degradation and provides guaranteed termination within the system.

\end{enumerate}

\subsection{Three-Tier Architecture} \label{sec31}
The proposed \hacn architecture is a three-tier system of connected components that deals with MAS consensus at various levels of scale and abstraction. 
\subsubsection{Local Consensus Clusters (Tier 1)}
The bottom level of the hierarchy organizes the agents into small groups of 3-5 members, depending on their professional expertise and task relevance. The K-means clustering is used to form the clusters. These clusters represent agentic knowledge and can be triggered whenever there is a change in task allocation. A capability matrix has been introduced to measure the agents' expertise by task relevance, the agent's reliability score, and historical accuracy information. Each cluster is equipped with an iterative confidence-weighted voting decision mechanism that allows a quick convergence on prediction accuracy.

\subsubsection{Inter-Cluster Coordination (Tier 2)}
This layer maintains inter-cluster consensus in terms of local clusters with authorized representatives. This tier enacts well-defined debate procedures, including dynamic timeouts and partial knowledge agreement, ensuring system-wide blocking and allowing cross-cluster knowledge integration. Further, adaptive threshold and confidence-voting strategies are used in this level to achieve the desired outcome.

\subsubsection{Global Orchestration Engine (Tier 3)}
The topmost layer performs system-wide coordination and ultimate arbitration on complex choices through a fallback approach when agents fail to settle at lower tiers. This engine keeps a history of the consensus and agent reliability values in global memory pools and adopts dynamic opportunity selection according to the tasks' complexity and time-interval limits.

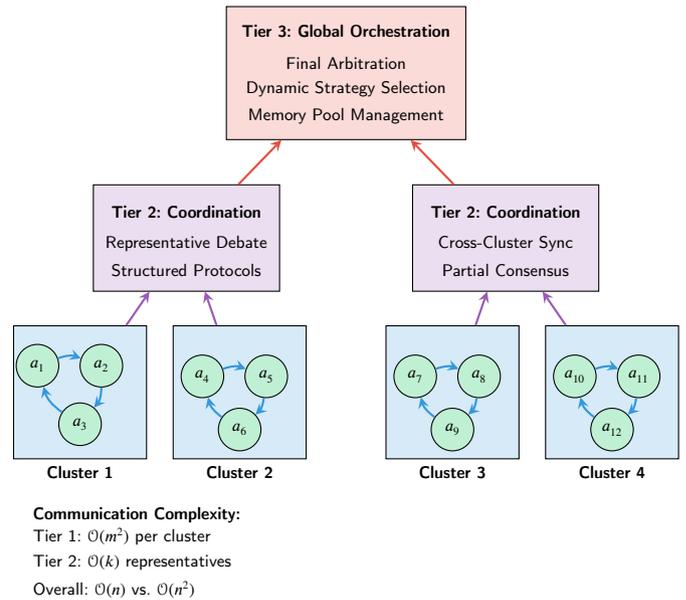
\begin{figure}[!htbp]
\centering
\begin{tikzpicture}[scale=0.7, transform shape]
    \tikzstyle{agent} = [circle, draw, fill=agentcolor!30, minimum size=8mm, inner sep=3pt]
    \tikzstyle{cluster} = [rounded rectangle, draw, fill=clustercolor!20, minimum width=15mm, minimum height=10mm]
    \tikzstyle{tier1} = [rectangle, draw, fill=tier1color!20, minimum width=25mm, minimum height=25mm]
    \tikzstyle{tier2} = [rectangle, draw, fill=tier2color!20, minimum width=35mm, minimum height=20mm]
    \tikzstyle{tier3} = [rectangle, draw, fill=tier3color!20, minimum width=45mm, minimum height=25mm]
    \tikzstyle{arrow} = [thick, ->, >=stealth]
    
    \node[tier3] (tier3) at (0,1) {};
    \node at (0,1.8) {\textbf{Tier 3: Global Orchestration}};
    \node at (0,1.2) {Final Arbitration};
    \node at (0,0.7) {Dynamic Strategy Selection};
    \node at (0,0.2) {Memory Pool Management};
    
    \node[tier2] (tier2-1) at (-3,-2.1) {};
    \node[tier2] (tier2-2) at (3,-2.1) {};
    
    \node at (-3,-1.6) {\textbf{Tier 2: Coordination}};
    \node at (-3,-2.2) {Representative Debate};
    \node at (-3,-2.7) {Structured Protocols};
    
    \node at (3,-1.6) {\textbf{Tier 2: Coordination}};
    \node at (3,-2.2) {Cross-Cluster Sync};
    \node at (3,-2.7) {Partial Consensus};
    
    \node[tier1] (cluster1) at (-5,-5) {};
    \node[tier1] (cluster2) at (-2,-5) {};
    \node[tier1] (cluster3) at (2,-5) {};
    \node[tier1] (cluster4) at (5,-5) {};
    
    \node[agent] (a1) at (-5.8,-4.5) {$a_1$};
    \node[agent] (a2) at (-4.6,-4.5) {$a_2$};
    \node[agent] (a3) at (-5,-5.6) {$a_3$};
    
    \node[agent] (a4) at (-2.7,-4.7) {$a_4$};
    \node[agent] (a5) at (-1.5,-4.7) {$a_5$};
    \node[agent] (a6) at (-2,-5.7) {$a_6$};
    
    \node[agent] (a7) at (1.3,-4.7) {$a_7$};
    \node[agent] (a8) at (2.5,-4.7) {$a_8$};
    \node[agent] (a9) at (2,-5.7) {$a_9$};
    
    \node[agent] (a10) at (4.3,-4.7) {$a_{10}$};
    \node[agent] (a11) at (5.5,-4.7) {$a_{11}$};
    \node[agent] (a12) at (5,-5.7) {$a_{12}$};
    
    \node at (-5,-6.5) {\textbf{Cluster 1}};
    \node at (-2,-6.5) {\textbf{Cluster 2}};
    \node at (2,-6.5) {\textbf{Cluster 3}};
    \node at (5,-6.5) {\textbf{Cluster 4}};
    
    \draw[arrow, tier3color] (tier2-1) -- (tier3);
    \draw[arrow, tier3color] (tier2-2) -- (tier3);
    
    \draw[arrow, tier2color] (cluster1) -- (tier2-1);
    \draw[arrow, tier2color] (cluster2) -- (tier2-1);
    \draw[arrow, tier2color] (cluster3) -- (tier2-2);
    \draw[arrow, tier2color] (cluster4) -- (tier2-2);
    
    \draw[arrow, tier1color, bend left=20] (a1) to (a2);
    \draw[arrow, tier1color, bend left=20] (a2) to (a3);
    \draw[arrow, tier1color, bend left=20] (a3) to (a1);
    
    \draw[arrow, tier1color, bend left=20] (a4) to (a5);
    \draw[arrow, tier1color, bend left=20] (a5) to (a6);
    \draw[arrow, tier1color, bend left=20] (a6) to (a4);
    
    \draw[arrow, tier1color, bend left=20] (a7) to (a8);
    \draw[arrow, tier1color, bend left=20] (a8) to (a9);
    \draw[arrow, tier1color, bend left=20] (a9) to (a7);
    
    \draw[arrow, tier1color, bend left=20] (a10) to (a11);
    \draw[arrow, tier1color, bend left=20] (a11) to (a12);
    \draw[arrow, tier1color, bend left=20] (a12) to (a10);
    
    \node[anchor=west] at (-6,-7.3) {\textbf{Communication Complexity:}};
    \node[anchor=west] at (-6,-7.7) {Tier 1: $\bigO(m^2)$ per cluster};
    \node[anchor=west] at (-6,-8.2) {Tier 2: $\bigO(k)$ representatives};
    \node[anchor=west] at (-6,-8.7) {Overall: $\bigO(n)$ vs. $\bigO(n^2)$};
\end{tikzpicture}
\caption{Proposed Three-Tier Architecture showing hierarchical consensus flow from local clusters (Tier 1) through inter-cluster coordination (Tier 2) to global orchestration (Tier 3)}
\label{fig:hacn_architecture}
\end{figure}

Figure \ref{fig:hacn_architecture} presents the suggested \hacn framework. The communication processes are streamlined in each tier, where the messages follow a logical format entailing solution contents, confidence ratings, justifications, and reliability markers of an agent. Such a multilevel strategy allows content-based and meta-cognitive evaluation of the consensus and keeps the communication load to a minimum. Thus, the agents improve their solutions from both intra-cluster and inter-cluster communication in each iteration and provide the best task awareness property.

\section{Proposed Hierarchical Adaptive Consensus Network} \label{sec4}

\subsection{Dynamic Cluster Formation}\label{sec41}
\begin{figure}[!ht]
\centering
\begin{tikzpicture}[scale=0.8, transform shape]
    \begin{axis}[
        width=8cm,
        height=6cm,
        xlabel={Expertise Domain 1},
        ylabel={Expertise Domain 2},
        grid=major,
        legend pos=outer north east,
    ]
    
    \addplot[only marks, mark=*, mark size=3pt, color=blue] coordinates {
        (0.2, 0.8) (0.3, 0.7) (0.1, 0.9) (0.25, 0.85)
    };
    \addlegendentry{Cluster 1}
    
    \addplot[only marks, mark=*, mark size=3pt, color=red] coordinates {
        (0.7, 0.3) (0.8, 0.2) (0.75, 0.35) (0.85, 0.25)
    };
    \addlegendentry{Cluster 2}
    
    \addplot[only marks, mark=*, mark size=3pt, color=green] coordinates {
        (0.5, 0.6) (0.6, 0.5) (0.45, 0.65) (0.55, 0.55)
    };
    \addlegendentry{Cluster 3}
    
    \addplot[only marks, mark=x, mark size=5pt, color=black, line width=2pt] coordinates {
        (0.21, 0.8) (0.77, 0.28) (0.53, 0.59)
    };
    \addlegendentry{Centroids}
    
    \end{axis}
\end{tikzpicture}
\caption{Dynamic agent clustering visualization showing agents grouped by capability similarity in a two-dimensional expertise space}
\label{fig:agent_clustering}
\end{figure}
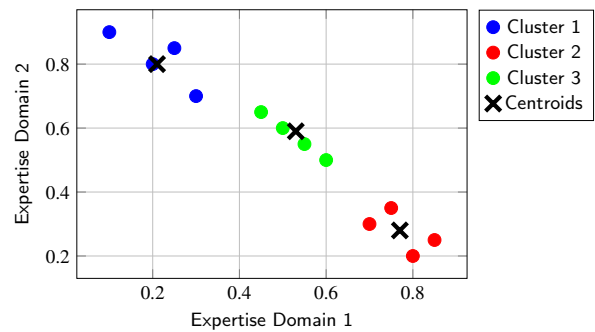

As discussed earlier, \hacn initially creates agent clusters dynamically to increase consensus convergence and solution quality. Figure \ref{fig:agent_clustering} illustrates the proposed dynamic agent clustering concept.
\begin{itemize}
\item K-means clustering with dynamic centroids is used to adapt to changes according to the number of jobs and available nodes. 
\item A square-root-based mathematical heuristic has been used to find out the optimal cluster number within the minimum and maximum cluster sizes of 3 and 5, respectively. 
\item The similarity measure integrates the semantic similarity of the agents' knowledge bases and correlation metrics of prior execution of analogous tasks for efficient clustering. 
\item The agents' relevance is estimated through a capability matrix, a multi-dimensional platform representing expertise domain, historical performance, and current task appropriation rating. 
\end{itemize}

The algorithm \ref{alg:cluster_formation} presents the proposed cluster formation mechanism. This approach has provisions to float a singleton cluster in case the agents cannot effectively be placed in a group. In addition, the clusters are reformed in real-time when the performance metrics reveal suboptimal clustering or the task requires other expertise, thus aiding in achieving efficient cluster communication. 

\begin{algorithm}[!ht]
\caption{Dynamic Cluster Formation}
\label{alg:cluster_formation}
\begin{algorithmic}
\Require Agent: $A = \{a_1, a_2, \ldots, a_n\}$, Task: $T$, History: $H$, $sz_{min} = 3$, $sz_{max} = 5$
\Ensure Cluster: $C = \{c_1, c_2, \ldots, c_k\}$
\State $n = length(A)$
\For{each agent $a_i \in A$}
    \State $v_i = f(\text{expertise}(a_i), \text{performance}(a_i, H), \text{relevance}(a_i, T))$ \Comment{\color{blue}$v_i$: capability vector}
\EndFor
\State initialize $k = \lfloor\sqrt{n}\rfloor$ \Comment{\color{blue}dynamic cluster count}
\Repeat
    \State centroids $=$ kmeansInit$(V, k)$ \Comment{\color{blue}$V$: capability matrix}
    \For{iter $= 1$ to $iter_{max}$}
        \For{each agent $a_i$}
            \State assign $a_i$ to closest centroid $c_j$
            \State apply size constraints [$sz_{min}, sz_{max}$]
        \EndFor
        \State update centroids based on assigned agents
        \If{convergence\_reached()}
            \State \textbf{break}
        \EndIf
    \EndFor
    \For{each cluster $c_j$}
        \State leader\_score $= \alpha \cdot \text{consensus\_success}(H) + \beta \cdot \text{expertise\_match}(T) + \gamma \cdot \text{availability}$
        \State select leader with highest score
    \EndFor
    \State evaluate cluster quality using silhouette coefficient
    \If{quality $<$ THRESHOLD}
        \State $k = k \pm 1$
    \EndIf
\Until{acceptable quality or max attempts reached}
\State \textbf{return} clusters $C$ with assigned leaders
\end{algorithmic}
\end{algorithm}

\subsection{Multi-Level Consensus Process} \label{sec42}
Consensus in \hacn is achieved after ensuring full coordination among all three levels of the decision-making process. Information and decisions are shared throughout the hierarchy based on the agent's needs and decision timing. Figure \ref{fig:consensus_process} depicts the proposed consensus process flow.

\begin{figure}[!ht]
\centering
\begin{tikzpicture}[scale=0.7, transform shape]
    \node[draw, rectangle, fill=tier1color!20, minimum width=3cm, minimum height=1.5cm] (local) at (0,0) 
          {\shortstack{Local Cluster\\Consensus}};
    
    \node[draw, rectangle, fill=tier2color!20, minimum width=3cm, minimum height=1.5cm] (inter) at (0,3) 
          {\shortstack{Inter-Cluster\\Coordination}};
    
    \node[draw, rectangle, fill=tier3color!20, minimum width=3cm, minimum height=1.5cm] (global) at (0,6) 
          {\shortstack{Global\\Orchestration}};
    
    \draw[thick, ->, >=stealth] (local) -- (inter) node[midway, right] {Escalation};
    \draw[thick, ->, >=stealth] (inter) -- (global) node[midway, right] {Final Decision};
    
    \node[draw, ellipse, fill=agentcolor!20] (conf1) at (-4,0) {$w_1 = c_1 \times h_1$};
    \node[draw, ellipse, fill=agentcolor!20] (conf2) at (-4,-1) {$w_2 = c_2 \times h_2$};
    \node[draw, ellipse, fill=agentcolor!20] (conf3) at (-4,-2) {$w_3 = c_3 \times h_3$};
    
    \draw[->] (conf1) -- (local);
    \draw[->] (conf2) -- (local);
    \draw[->] (conf3) -- (local);
    
    \node[draw, diamond, fill=yellow!20] (thresh) at (4,1.5) 
          {\shortstack{Adaptive\\Threshold\\$\theta(T, \tau)$}};
    
    \draw[->] (thresh) -- (local);
    \draw[->] (thresh) -- (inter);
    
    \node[draw, rectangle, fill=orange!20] (time) at (4,4.5) 
          {\shortstack{Time\\Constraint\\$\tau$}};
    
    \draw[->] (time) -- (thresh);
    \draw[->] (time) -- (global);
    
    \node[anchor=west] at (-5,-4) {\textbf{Notation:}};
    \node[anchor=west] at (-5,-4.5) {$c_i$: Agent confidence, $h_i$: Historical accuracy, $\tau$: Time limit};
    \node[anchor=west] at (-5,-5) {$w_i$: Vote weight, $\theta$: Consensus threshold};
\end{tikzpicture}
\caption{Multi-level consensus process flow showing hierarchical escalation with confidence-weighted voting and adaptive thresholds}
\label{fig:consensus_process}
\end{figure}

\begin{itemize}
    \item As stated earlier, the Tier-1 establishes intra-cluster communication among agents based on task-relevant prompts and expertise. 
    \item Tier-2 balances the consensus of several local clusters with similarities to find the best solution, coupled with confidence edges and an adaptive threshold. This meta-clustering identifies the general agreement zone and underlines particular areas of disagreement that need special debate. Later, the Structured debate protocols are used in the coordination process. This procedure goes on until there is a mutual understanding or the time is up.
    \item Tier-3 solves the inconsistencies found on the lower levels of the consensus process. The fallback feature incorporated here provides system reliability in such poor conditions. The system switches to majority voting among the cluster representatives when the technical assessment fails to separate the solutions. In case of a vote being tied, predetermined tie-breakers are applied depending on the complexity of the solution, resources needed, and historical performance of that cluster.
\end{itemize}

Algorithm \ref{alg:hgca} presents the pseudocode of how agentic consensus is achieved in the proposed \hacn framework. After the Tier-1 forms the local agent clusters, the consensus is iteratively established through integrating inter-cluster communication and global arbitration. The algorithm is segregated into four phases of structured debate, similarity-based clustering, feasibility validation, and updation of the agent reliability matrix. 

The notations used in the algorithm are presented below.
\textbf{Notations:} Cluster Solution: $CS$, Task: $T$, Global Memory: $GM$, Time\_limit: $\tau$, Final Decision: $FD$, Updated Metrics: $M$, Current time: $Time_{curr}$, Representative: $repr$, Cluster component: $cs^{clus}$, Cluster consensus strength: $cs^{str}$, Cluster solution: $cs^{sol}$, Arguments: $args$, Opposing solution: $sol_{opp}$, Updated representative: $repr_{upd}$, Counter argument: $args_{c}$, Refined solution: $sol_{ref}$, Similarity matrix: $mat_{sim}$, Threshold: $\Upsilon$, Candidate solution: $sol_{cand}$, Cluster representative: $repr_c$, Cumulative score: $sc$, Feasible solutions: $sol_{feas}$, Feasibility check: $check_{feas}$, Adjusted score: $score_{adj}$

\begin{algorithm}[!ht]
\caption{Proposed Hierarchical Global Consensus Arbitration (HGCA)}
\label{alg:hgca}
\begin{algorithmic}
\Require $CS$, $T$, $GM$, $\tau$
\Ensure $FD$, $M$\\
\State \textbf{begin procedure}
\State $start\_time \leftarrow Time_{curr}()$ \Comment{\color{blue}Initialize execution timer}
\phase{Inter-cluster coordination via representative debate}
\State $repr \leftarrow \{\}$ \Comment{\color{blue}Initialize cluster representative}
\For{each $cs_i \in CS$} \Comment{\color{blue}Select representative}
    \State $rep \leftarrow$ selectRepr ($cs_i^{clus}, cs_i^{str}$)  
    \State $repr \leftarrow repr \cup \{\langle rep, cs_i^{sol} \rangle\}$  
\EndFor
\State $debate\_solutions \leftarrow \{\}$ \Comment{\color{blue}Do debate solution process}
\State $debate\_rounds \leftarrow$ calcDebateRound ($|repr|$, $\tau$)
\For{$round \leftarrow 1$ \textbf{to} $debate\_rounds$}
    \If{$Time_{curr}() - start\_time > 0.7 \times \tau$}
        \State \textbf{break} \Comment{\color{blue}Reserve time for arbitration}
    \EndIf
    \State $args \leftarrow \{\}$  \Comment{\color{blue}args: arguments}
    \For{each $\langle rep_i, sol_i \rangle \in repr$}
    \LComment{\color{blue}perform Set Minus operation ($\setminus$)}
        \State $sol_{opp} \leftarrow repr \setminus \{\langle rep_i, sol_i \rangle\}$ 
        \State $arg \leftarrow$ generateArg ($rep_i, sol_i, sol_{opp}, GM$)
        \State $args[sol_i] \leftarrow arg$
    \EndFor
    \LComment{\color{blue}Cross-evaluate arguments and update solutions}
    \State $repr_{upd} \leftarrow \{\}$
    \For{each $\langle rep_i, sol_i \rangle \in repr$}
    \LComment{\color{blue}perform Set Minus operation ($\setminus$)}
        \State $args_c \leftarrow args \setminus \{args[sol_i]\}$
        \LComment{\color{blue}Incorporate counter arguments}
        \State $sol_{ref} \leftarrow$ counterArg ($sol_i$, $args_c$) 
        \State $repr_{upd} \leftarrow repr_{upd} \cup \{\langle rep_i, sol_{ref} \rangle\}$
    \EndFor
    \State $repr \leftarrow repr_{upd}$
\EndFor
\LComment{\color{blue}Extract final debate solutions}
\For{each $\langle rep_i, sol_i \rangle \in repr$}
    \State $debate\_solutions \leftarrow debate\_solutions \cup \{sol_i\}$
\EndFor
\algstore{myalg}
\end{algorithmic}
\end{algorithm}
\begin{algorithm}
\addtocounter{algorithm}{-1}
\begin{algorithmic}
\algrestore{myalg}
\phase{Similarity-based solution clustering}
\LComment{\color{blue}Compute Similarity Matrix}
\State $mat_{sim} \leftarrow$ compSimMat ($debate\_solutions$)
\LComment{\color{blue}Do hierarchical clustering}
\State $cs^{sol} \leftarrow$ hierCluster ($mat_{sim}, \Upsilon =0.7$) 
\State $sol_{cand} \leftarrow \{\}$ \Comment{\color{blue}Find candidate solution}
\For{each $cs_i^{sol} \in cs^{sol}$}
    \LComment{\color{blue}Select representative solution from each cluster}
    \State $repr_c \leftarrow$ selectClusterRepr ($cs_i^{sol}$, $mat_{sim}$)
    \LComment{\color{blue}Compute Cumulative Similarity}
    \State $cs_i \leftarrow$ compSim ($repr_c$, $debate\_solutions$)
    \State $sol_{cand} \leftarrow sol_{cand} \cup \{\langle repr_c, cs_i \rangle\}$
\EndFor
\phase{Global arbitration with feasibility validation}
\LComment{\color{blue}Find Feasible solutions}
\State $sol_{feas} \leftarrow \{\}$
\LComment{\color{blue}Perform Technical feasibility ($\Omega$), consistency ($\Phi$) and resource constraint ($\Psi$) checks using prior decisions stored in Global Memory ($GM.pd$) and Task Resources ($T_{res}$)}
\For{each $\langle sol_i, score_i \rangle \in sol_{cand}$}
    \State $check_{feas} \leftarrow \{\Omega(sol_i, T), \Phi(sol_i, GM.pd),\\ \Psi(sol_i, T.res)\}$
    \If{allCheckPass ($check_{feas}$)}
    \LComment{\color{blue}Add a feasibility bonus}
        \State $score_{adj} \leftarrow score_i \times$ feasBonus ($check_{feas}$)
        \State $sol_{feas} \leftarrow sol_{feas} \cup \{\langle sol_i, score_{adj} \rangle\}$
    \EndIf
\EndFor
\LComment{\color{blue}Final decision selection}
\If{$|sol_{feas}| > 0$}
    \State $FD \leftarrow$ argmax($sol_{feas}$, key$=score_{adj}$)
\Else
    \LComment{\color{blue}Fallback to majority voting among original cluster solutions}
    \State $vc \leftarrow \{\}$  \Comment{\color{blue}vote count: $vc$}
    \For{each $cs_i \in CS$}
        \State $vc[cs_i^{sol}] \leftarrow cs_i^{str} \times |cs_i^{clus}|$
    \EndFor
    \State $FD \leftarrow$ argmax($vc$)
\EndIf
\phase{Update agent reliability metrics}
\State $M \leftarrow$ updateReliability ($CS$, $FD$, $GM$)
\LComment{\color{blue}Update global memory $GM$ using consensus history ($GM.ch$) and agent metrics ($GM.am$)}
\State $GM.ch \leftarrow GM.ch \cup \{\langle T, CS, FD, Time_{curr}() \rangle\}$
\State $GM.am \leftarrow$ mergeMetrics ($GM.am$, $M$)
\State \textbf{return} $\langle FD, M \rangle$
\State\textbf{end procedure}
\end{algorithmic}

\end{algorithm}

\subsection{Adaptive Decision Rules}\label{sec43}
\hacn introduces complex decision rules to keep consistency (homogeneity) and fairness distinct in various cases of consensus. These rules keep changing across diverse operational environments and are used during threshold adjustment, partial consensus, and time-bound decision-making. 

As discussed earlier, consensus thresholds are context-dependent. They help adapt to the required agreement level depending on the importance of the task, time availability, the agent's reliability during job execution, and past success on similar undertakings. Here, we set the threshold values as per an entropy-based computation of the number of available solution attempts and their quality.

Moreover, the partial consensus acceptance allows \hacn to move to the stage of a clear majority agreement among the agents. Further, the minority views are stored in a position that may affect future decisions and provide feedback on the solution. The Time-bounded decision-making feature incorporates gradual urgency measures for reaching a consensus and maintaining decision quality during emergencies. We have exploited the cached advice on analogous past choices and a simplified voting algorithm to achieve our objective.

\section{Evaluation Methodology}\label{sec5}
A thorough assessment of \hacn needs to have both theoretical confirmation as well as empirical testing under a wide range of operational conditions. We have used simulation, comparative benchmarking, and deployment case study as evaluation tools.

\subsection{Simulation Environment Setup}  \label{sec51}
A large-scale multi-agent simulation environment is set to capture realistic task mixing distributions, agent capability distributions, and network conditions. We have taken a range of 10 to 1000 agents to present the task complexity settings and variety of failure conditions. Furthermore, the simulation provides extensive information on agents' reliability distribution and their abilities over four performance metrics.
\begin{enumerate}
    \item Communication overhead in terms of messages and communications,
    \item Convergence time (measured in seconds) from the beginning of task execution till the end,
    \item Number of clusters formed, and
    \item Number of agents per cluster.
\end{enumerate}

\subsection{Result Analysis} \label{sec52}
\begin{figure*}[!htbp]
\centerline{\includegraphics[width=0.9\textwidth]{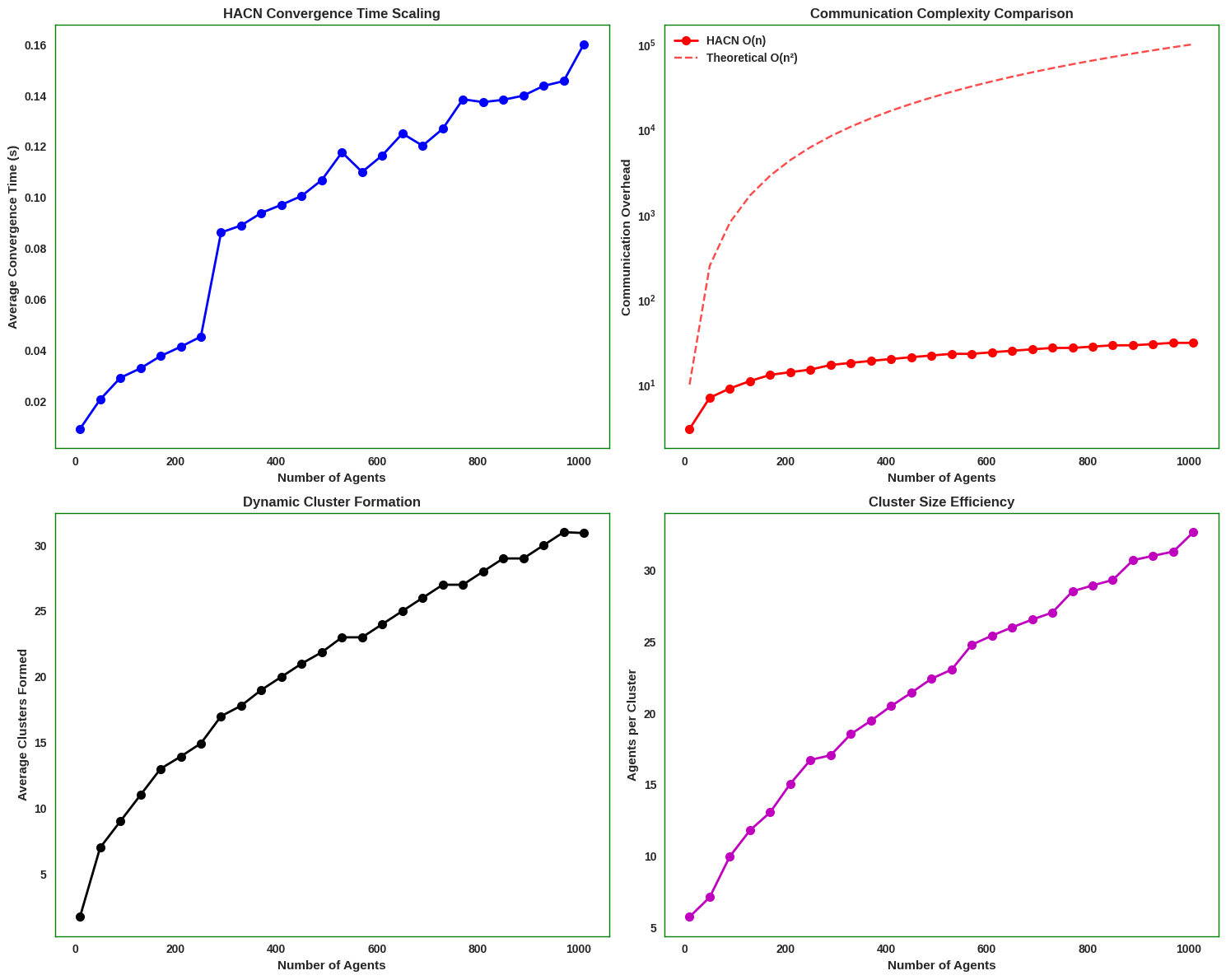}}
\caption{Performance of Proposed \hacn Architecture}
\label{fig_hacn}
\end{figure*}

Figure \ref{fig_hacn} provides the \hacn performance on the aforementioned metrics. It is evident from the figure that the \hacn produces a uniform convergence rate with increasing number of agents. The model is taking 0.045 seconds at 250 agents, in contrast to 0.086 seconds at 290 agents. This difference is mainly caused by the number of clusters formed, as seen under the dynamic cluster formation subplot.

Further, the communication overhead of \hacn is significantly lower than the existing baseline model, which has $\bigO(n^2)$ complexity. As we approach 1000 agents, we can see that \hacn exhibits an almost linear plot with around 310 communication messages. Additionally, the plot presenting cluster information shows that \hacn achieved 31 clusters with 1000 agents grouped by around 32 agents per cluster.
\subsection{Comparative Baseline} \label{sec53}
\begin{figure*}[!htbp]
\centerline{\includegraphics[width=0.9\textwidth]{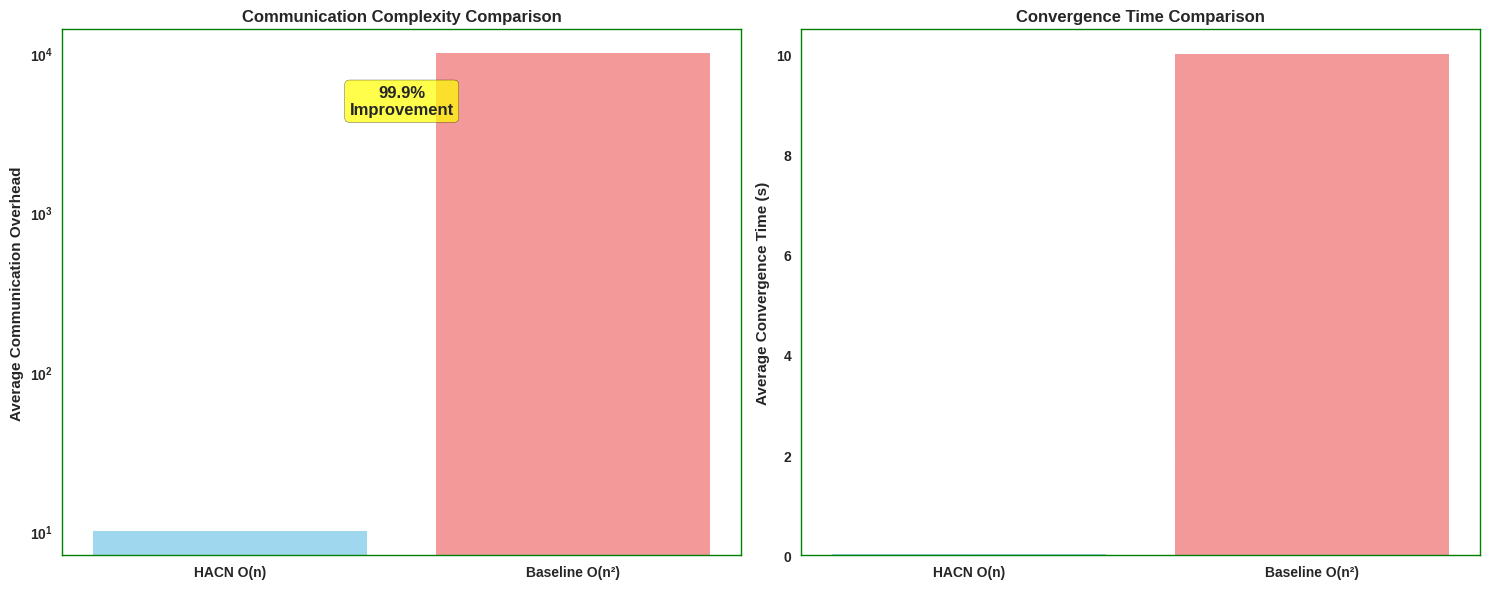}}
\caption{Comparison between \hacn and existing consensus frameworks}
\label{fig_comp}
\end{figure*}

A comparative experimentation with a baseline consensus framework having $\bigO(n^2)$ complexity has been presented in Figure \ref{fig_comp} in terms of average communication overhead and convergence time. The comparison is done on 100 agents handling 25 different tasks.

The comparison showed communication overhead reduction of 99.9\% with \hacn producing average communication of 10 as opposed to $10^4$
under the baseline model. Similarly, \hacn takes 0.034 seconds to reach a consensus with convergence in contrast to 10 seconds of baseline time.

\section{Implementation Considerations}\label{sec6}
The modular structure of \hacn allows its selective implementation depending on operation limitations and demands. However, a lot of consideration is needed on system architecture, memory management, fault tolerance, and performance optimization when implementing in the real world.

\subsection{Agent Deployment Architecture}  
\hacn provides various deployment patterns, such as containerized microservices, distributed computing frameworks, and cloud-native architecture. A specific agent can act as a stand-alone service that shares standardized APIs to communicate with other clusters and can also perform global coordination.

The deployment architecture contains load-balancing nodes during cluster formation services, distributed message queuing devices during inter-cluster communication, and redundant orchestration engines for handling exceptions. Dynamic agent registration and cluster reformation are also changeable through service discovery mechanisms.

\subsection{Memory Management for Consensus History} \hacn adopts a distributed memory management for local agent caches and cluster memory pools for facilitating both intra-cluster and inter-cluster communication. Takeovers of consensus history are done with optional retention policies to provide a trade-off between the quantity of accessible historical contexts against storage requirements. Further, memory optimization involves the use of compression routines to encode previous messaging, intelligent pre-caching of commonly used decision patterns, and garbage collection of stale decision consensus.

\subsection{Network Topology Optimization} \hacn can fit into both fully-connected, hierarchical, and mesh network topologies, because the communication latency and bandwidth limits are taken into account during cluster formation. The Quality of Service (QoS) mechanism hierarchically takes precedence over non-essential consensus messages during a congested network, thus maintaining system responsiveness. Moreover, the framework has automatic network partitioning detection and graceful degradation processes to ensure functionality when there is a connectivity failure.

\subsection{Fault Tolerance and Recovery Mechanisms} \hacn has a complete fault tolerance mechanism to detect failures in its three-tiered architecture. The Byzantine fault tolerance system guards against an involuntary or polluted agent that might be trying to interfere with the consensus procedures. Furthermore, state restoration is done according to the checkpoints implemented during model construction. 

\section{Convergence and Scalability Analysis}\label{sec7}
This section provides a theoretical analysis of achieving agentic consensus convergence and scalability as the number of agents increases.
\subsection{Convergence Properties}
\hacn confirms guaranteed convergence through hierarchical escalation and fewer iterations. Initially, the local convergence is achieved during intra-cluster formation in Tier-1 as per Eq. \eqref{eq1}.
\begin{equation} \label{eq1}
    P(\text{convergence}) \geq 1 - (1 - p)^k
\end{equation}
where $p$ is the probability of a match in any iteration, and $k$ is the upper limit of iterations. This ensures attainment of a high consensus probability when enough iterations are done. 

Furthermore, formal analysis has shown that the convergence probability of \hacn approaches one and the convergence time becomes a logarithmic function of the system size as the number of escalation levels increases. Additionally, \hacn converges with more than a probability of 0.95 under realistic assumptions on the agent's reliability distributions and task variations. These limits are not sensitive to scale differences, implying high performance stability. 

\subsection{Scalability Improvements}
\hacn enjoys significant scalability benefits over other flat-consensus algorithms with improved communication complexity at $\bigO(n)$ and reduced network cost and computational overheads. It has been mathematically established that our model, consisting of $n$ agents arranged in $c$ clusters, has $\bigO(n + k^2)$ communication complexity, where $k=\frac{n}{c}$ represents the cluster number. This is a huge improvement over traditional methods as $k \ll n$.

\subsubsection{Load Balancing Through Hierarchical Decomposition} .  Computational load is naturally distributed throughout the three-tier architecture, leading to no bottlenecks. Solutions are developed at local clusters, whereas coordination and arbitration are done at higher levels. This decomposition allows local clusters to operate in parallel and large clusters on each decision domain. Finally, a consensus is reached with efficient resource utilization with less time consumption.

\subsubsection{Resource Utilization Efficiency Gains} Empirical study suggests that \hacn reduces the overall communications overhead by $99.9\%$ over existing full-connectivity models. As discussed, the intra-cluster formation in Tier-1 and inter-cluster communication in Tier-2 provide a balanced load on the available computational resources. Such efficiencies are easier to scale in higher-order systems. Hence, we can say that \hacn is well-suited to large-scale deployment.

\section{Conclusion and Future Work}\label{sec8}
The proposed \hacn framework overcomes severe shortcomings of consensus problems present in the existing collaborative multi-agent AI systems. With the three-tier architecture using local clustering, inter-cluster coordination, and global orchestration functions, \hacn enjoys great enhancement in scalability, adaptability, and convergence reliability. The framework alleviates the communication complexity of agreement to $\bigO(n)$ and guarantees optimal consensus quality by a hierarchical decomposition of the original problem. Further, the adaptive decision rules allow for attaining the optimum performance in various conditions without requiring manual reconfiguration. The confidence-weighted voting mechanism used in \hacn guarantees controllable decision-making in heterogeneous capabilities of agents. 

Experimental analysis proves convergence with hierarchical escalation procedures and probabilistic convergence error bounds being well over 99\% under realistic assumptions. Further, the logarithmic scale in the number of agents also establishes our model's true large-scale nature. \hacn allows deployment of large-scale multi-agent systems in real-world scenarios where other methods fail due to scalability issues. With modular architecture, \hacn can selectively be implemented on a per-use basis in a variety of contexts, ranging from distributed computing, robotics, and even social networks. 

However, the present constraints are computational overhead on dynamic clustering and possible performance loss due to the possibility of extreme network partitioning. The future research includes machine learning to supplement the predictive cluster creation, a blockchain-based trust model to validate the consensus, support for an asynchronous environment, and the deep trust model based on reputation systems and behavioral analysis. The current design is a major step forward in practical, scalable, and reliable consensus protocols to allow large-scale collaborative deployments of AI. The design principles of hierarchical decomposition, adaptive decision-making, and confidence-weighted voting can form the basis of future efforts in scaling distributed artificial intelligence and autonomous system coordination required of critical AI applications.

\end{document}